\documentclass{ws-ijgmmp}
\usepackage{cite}
\usepackage{subfigure}
\usepackage{color}

\begin{document}

%
\catchline{}{}{}{}{}
%

\title{Observational constraints on holography in $(2 + 1)$-dimensional cosmology with a generalized equation of state}

\author{Praveen Kumar Dhankar}
\address{Symbiosis Institute of Technology, Nagpur Campus, Symbiosis International (Deemed University), Pune 440008, Maharashtra, India\\ 
\email{pkumar6743@gmail.com}}

\author{Safiqul Islam}
\address{Department of Basic Sciences, General Administration of Preparatory Year, King Faisal University, P.O. Box 400, Al Ahsa 31982, Saudi Arabia \& Department of Mathematics and Statistics, College of Science, King Faisal University, P.O. Box 400, Al Ahsa 31982, Saudi Arabia\\
\email{sislam@kfu.edu.sa}}

\author{Saibal Ray}
\address{Centre for Cosmology, Astrophysics and Space Science (CCASS), GLA University, Mathura 281406, Uttar Pradesh, India\\ 
\email{saibal.ray@gla.ac.in}}

\author{Aritra Sanyal}
\address{Department of Mathematics, Jadavpur University, Kolkata, West Bengal 700032, India\\
\email{aritrasanyal1@gmail.com}}

\author{S. K. Maurya}
\address{Department of Mathematical and Physical Sciences, College of Arts and Sciences, University of Nizwa, Nizwa 616, Sultanate of Oman\\
\email{sunil@unizwa.edu.om}}

\maketitle	

\abstract{In this study we explore the cosmic holographic principle, as proposed by Fischler and Susskind~\cite{Fischler}, within the framework of $(2 + 1)$-dimensional cosmological models. A generalized equation of state is employed, given by $p = (\zeta - 1)(\rho + \rho_0)$, where $\zeta$ and $\rho_0$ are treated as two free parameters. The analysis confirms the validity of the holographic principle in all flat and open universes. However, for a $(2 + 1)$-dimensional closed universe, we apply the method proposed by Kaloper and Linde~\cite{Kaloper}, and observe that the holographic principle is generally not satisfied. Furthermore, we examine the stability of the proposed model using the Markov chain Monte Carlo (MCMC) method, and estimate the best-fit values for the model parameters based on observational Hubble data sets.}


\section{Introduction}\label{intro}

Literature reveals that much significant attention has been devoted to exploring the gravitational theories in dimensions other than four. The reason is driven by various motivations, however the primarily one comes from string theory, grand unified theories, and quantum gravity. In particular, general relativity in three spacetime dimensions exhibits a number of special and simplifying features. Notably, in $(2 + 1)$-dimensions, there are no gravitational waves or black hole solutions in the absence of a negative cosmological constant. Additionally, the Weyl curvature tensor vanishes identically, and the weak-field limit does not correspond to Newtonian gravity.

Numerous studies have examined the structure of general relativistic gravitational theory in $(2 + 1)$-dimensional spacetime~\cite{Giddings}--\cite{Barrow1}. The unique characteristics of Einstein's field equations in two spatial and one temporal dimension is the principle motivation for focusing on $(2 + 1)$-D. Just as lower-dimensional models are frequently applied in quantum field theory to gain physical insight, it is similarly useful to explore the gravitational models in lower dimensions. This approach is not only theoretically well-founded but also potentially illuminating. It is widely believed that studying $(2 + 1)$-dimensional gravity can provide valuable insights into the more physically relevant $(3 + 1)$-dimensional theory~\cite{Garica}.

Over the past decade, several studies on extended gravity theories have been done~\cite{Meng}, which aim to explore the observed cosmic acceleration. Such approaches usually involve modifying the equation of state, $p = \omega\rho$, where $\omega < 0$ or introducing dark energy. These models are however hydro-dynamically unstable. To overcome this issue, a more general linear equation of state, $p = \alpha(\rho - \rho_0)$, was proposed by~\cite{Babichev}, where $\alpha$ and $\rho_0$ are constants. This non-homogeneous linear equation of state can describe both hydro-dynamically stable fluids ($\alpha > 0$) and unstable ones ($\alpha < 0$). In the special case when $\rho_0 = 0$, the model reduces to the linear perfect fluid with a constant $\omega$.

It is well-known in black hole theory that the total entropy of matter inside a black hole is proportional to the area $A$ of its event horizon, given by the relation $S/A=1/4$. This idea suggests that, under certain conditions, all information about a physical system is encoded on its boundary, implying that the entropy does not exceed the boundary area in Planck units. This concept is known as the holographic principle~\cite{Hooft, Susskind}. The principle goal of the holographic principle is to extend this idea to a broader class of physical scenarios. In its most radical form, it implies that our universe is effectively two-dimensional, as all the information about physical processes is stored on its boundary. This conjecture is particularly intriguing, and the physical implications of its most radical formulation could be profound~\cite{Kaloper}. A notable generalization of this holographic principle to cosmology was first proposed by~\cite{Fischler}. A key aspect of their work is that the principle holds for flat and open universes as well, provided that the equation of state satisfies the condition $0 \leq p \leq \rho$. However, in the case of a closed universe, this formulation is violated. Rama~\cite{Rama} later argued that the principle can still be preserved in closed universes if the universe contains exotic matter with negative pressure.

The issue becomes more pronounced when one examines a universe with a negative cosmological constant~\cite{Kaloper}. In such scenarios, the holographic principle is no longer valid, regardless of whether the universe is closed, open, or flat. 

Holographic considerations have been used in the investigation of the pre-big bang scenario~\cite{Bak}-\cite{Veneziano}. Based on these studies, the theory was claimed to resolve the entropy problem inherent in the pre-big bang theory~\cite{Veneziano}. However, this conclusion appears to be in conflict with the results of~\cite{Kaloper1}.

Motivated by the work of Meng et al.~\cite{Meng1}, in this paper we examine the fundamental assumptions of cosmic holography within $(2+1)$-dimensional cosmological models with a generalized equation of state. The mathematical simplicity of these models may provide further insight into the underlying physics. We aim to explore the core assumptions of cosmic holography and demonstrate that the holographic principle holds for open and flat universes with a general equation of state. However, by applying the method proposed by~\cite{Kaloper, Wang}, it is found that the harmonic principle in a $(2 + 1)$-dimensional closed universe cannot be manifested for $\lambda<0$ where $\lambda = 2(\gamma-1)^{2}/\gamma$. Furthermore, $\lambda$ serves as the negative cosmological constant when $\rho_0<0$. The presence of a negative cosmological constant provides further motivation to seek a revised version of the cosmological harmonic principle.

For the observational datasets, we have selected the data set $H(z)$, which consists of 30 measurements, to obtain the best-fit values for the model parameters. We employ a Monte Carlo Markov chain (MCMC) analysis to determine the model parameters using the most recent data. Observational data analysis is carried out using the $\chi^2$-minimization technique, focusing on the $H(z)$ dataset.

The structure of the paper is envisaged as follows: Section 2 presents the basic equations for the FRW universe in the context of a $(2+1)$-dimensional spacetime. In Section 3, solutions to the Einstein field equations are derived under different curvature conditions $k= -1, 0, +1$, for various universes, including a dust-filled universe and a radiation-dominated universe. Section 4 examines the condition for a flat universe with $\lambda<0$. In Section 5, the model parameters are constrained using the Hubble dataset, and the best-fit values are obtained through the implementation of the MCMC method with observational data. Justification of the Figures are given in Section 6. Concluding remarks are provided in Section 7.

\section{FLRW Model and Friedmann equations in (2+1)-dimensional spacetime}

~~~We consider the (2+1)-dimensional Friedmann-Lemaître-Robertson-Walker (FLRW) line element~\cite{Khadekar} as
\begin{equation}
	\label{eq1} ds^{2} = dt^{2}-a^{2}(t)\left [\frac{dr^{2}}{1-k r^{2}}+ r^{2}d\theta^{2}\right],
\end{equation}
where $a(t)$ means the scale factor. The coordinates ($t$, $r$, $\theta$) denotes the co-moving coordinates and the constant $k$ denotes the curvature of the space $k=0,1,-1$ for flat, closed and open universe respectively.

The Einstein field equations in $(2+1)$-dimension spacetime have been proposed in~\cite{Cornish}
\begin{equation}
\label{eq2} G_{ij} = R_{ij}-\frac{1}{2}R g_{ij} = {2}\pi G T_{ij}.
\end{equation}

The energy momentum tensor with matter in $(2+1)$-dimensional spacetime is given by 
\begin{equation}
	\label{eq3} T_{ij} = (\rho + p) u_{i} u_{j} - p g_{ij},
\end{equation}
where $u^{i}$ is the velocity of three vector with $u^{i}u_{j}=-1$. 

The Einstein field equation (2) with the help of line element (1) in (2+1)-dimensions are given as~\cite{Cornish}
\begin{equation}
	\label{eq4} \left(\frac{\dot{a}}{a}\right)^{2} + \frac{k}{a^{2}}={2}\pi G \rho,
\end{equation}

\begin{equation}
	\label{eq5} \frac{\ddot{a}}{a}=-{2}\pi G \rho,
\end{equation}
where over dot means the derivative with respect to cosmic time $t$.

The energy-momentum conservation equation in (2+1)-dimensional spacetime is~\cite{Khadekar}
\begin{equation}
\label{eq6} \dot{\rho}+2 H (\rho+p)=0,
\end{equation}
where $H=\frac{\dot{a}}{a} $ is the Hubble parameter, $\rho$ and $p$ are the energy density and pressure.

\section{Solution to the field equations}

Let us assume the generalized equation of state of the form as
\begin{equation}
\label{eq7}  p=(\zeta-1)(\rho+\rho_0),
\end{equation}
where $\zeta$ and $\rho_0$ are two parameters. For the dust filled universe $\zeta= 1$, i.e. $p = 0$. 

 The energy conservation Eq. (6) becomes
\begin{equation}
\label{eq8} \rho a^{2}= \text{const} = {d_0}{a_0}^{2}.
\end{equation} 

From Eq. (3), we get
\begin{equation}
\label{eq9} \dot{a}^{2}=2 \beta_0 G-k,
\end{equation}
where $\beta_0= \pi d_0 {a_0}^{2}$. 

This equation requires $\beta_0 \geq k/2G$ and hence the solution of Eq. (9) gives
\begin{equation}
\label{eq10} a(t) = a_0 + \sqrt{2 G \beta_0 - k}t.
\end{equation} 

This equation indicates that the universe is filled with dust and perpetually expands regardless of the value of $k$. The analogous solutions have already been addressed by~\cite{Wang}. The particle horizon is defined as the distance traversed by the light cone at the singularity $t = 0$. Hence, one may define 
\begin{equation}
\label{eq11} L_H = a(t){\int_{t= 0}^{t= t_0}\frac{dt'}{a(t')}}=a(t) r_{H(t)},
\end{equation}
where $r_H $ is the comoving distance to the horizon.

\subsection{Dust filled universe, i.e. $p=0$ for $\zeta= 1$}

The comoving horizon from Eq. (10) is
\begin{equation}
\label{eq12} r_H = {\frac{1}{\sqrt{2G \beta_0-k}}} {\log\left(\frac{a(t)}{a_0}\right)}.
\end{equation}

The total entropy inside the horizon divided by the area grows in (2+1)-dimensions is given \cite{Wang}
\begin{equation}
\label{eq13} \frac{\mathbb{S}}{\mathbb{A}} = \sigma \frac{ L_H}{a^{2}} = \sigma \frac{ r_H}{a}=\frac{\sigma}{\sqrt{2G \beta_0-k}}\left(\frac{a(t)}{a_0}\right),
\end{equation}
where $\sigma$ represents the comoving entropy density and satisfies $\sigma{a_0}^2$ at the initial stage. Consequently, in the context of an expanding universe, the constraint will be fulfilled in the cases of both flat and open universes within a (2+1)-dimensional framework. This result is examined from a different perspective by~\cite{Wang}.

\subsection{Radiation dominated universe, i.e. $\rho = 2p$}

For this particular case, let us assume that $\rho_0=0$ and $\zeta = 3/2$, then the conservation Eq. (6) becomes 
\begin{equation}
\label{eq14} \rho a^{2} = \text{const} = d_0 {a_0}^{3}.
\end{equation} 

The field Eqs. (4) and (5) can be rewritten as 
\begin{equation}
\label{eq15} \dot{a}^2=\frac{2G \beta_0}{a}-k,
\end{equation}

\begin{equation}
\label{eq16} \ddot{a}=-\frac{\beta_0 G}{a^2}.
\end{equation}

\subsection{When k = 0, i.e. flat universe}

For the flat universe the solution of Eq. (15) can be expressed as~\cite{Wang}
\begin{equation}
\label{eq17}  a(t)=\left[\frac{3}{2} (2G\beta_0)^{1/2}\right]^\frac{2}{3} t^{2/3}.
\end{equation}

The comoving horizon $r_H$ is given as
\begin{equation}
\label{eq18} r_H=\frac{3}{\left[\frac{3}{2} (2 G \beta_0)^{1/2}\right]^{2/3}} t^{1/3}.
\end{equation}

The comoving total entropy area ratio behave as given
\begin{equation}
\label{eq19}  \frac{\mathbb{S}}{\mathbb{A}} \sim t^{-1/3}.
\end{equation}

The ratio $``{\mathbb{S}}/{\mathbb{A}}"$ does not increase with respect to time ans hence, for the expanding universe holographic bound will be satisfied. The entropy density at the Planck time $t = 1$ could not be greater than $1$, so one may say that initially ${\mathbb{S}}/{\mathbb{A}} = \sigma \leq 1$. In an expanding universe the number of degree of freedom associated with the cosmological horizon, or with apparent horizon, or with horizon of a would be back hole which provides holographic constraints on entropy, rapidly change in time. This results is in agreement with that in (3+1)-dimensional spacetime. The behavior of the open universe, i.e. $k =-1$, is similar to that of flat universe.

\subsection{When $k = 1$, i.e. closed universe }

In a closed universe, the initial horizon area is infinitesimally small, then expands to a maximum, and ultimately vanishes. The degrees of freedom associated with such a surface are significantly influenced by time, even in an adiabatic evolution of the cosmos, whereas the overall degrees of freedom in the bulk remain conserved~\cite{Bousso,Bousso1}. During the radiation-dominated phase, the holographic bound is breached at the turning point, indicating that the violations occur much prior to the universe attaining the future Planck era~\cite{Wang1}.

\section{A special case for flat universe with $\lambda < 0$}

Let us discuss the solutions of the field equations for the flat cosmological models. We examine the holographic principle with generalized equation of state with the assumptions that $\lambda<0$ for $\rho_0<0$.  

From the conservation Eq (6) with the generalized equation of state Eq. (7), we have
\begin{equation}
\label{eq20} \rho = -B_0 + \frac{\rho_0}{a^{2 \zeta}}.
\end{equation}

In this case the field Eqs. (4) and (5) can be written as
\begin{equation}
\label{eqn21} \left(\frac{\dot{a}}{a}\right)^{2} = 2\pi G \left(-B_0 +\frac{\rho_0}{a^{2\zeta} } - \frac{k}{a^{2}} \right),
\end{equation}

\begin{equation}
	\label{eqn22} \ddot{a} = -{2} \pi G (\zeta - 1) \left(-{B_0}(\zeta - 1) +\frac{\rho_0}{a^{2 \zeta}}  \right) a.
\end{equation}

For $k = 0$, Eq. (21) becomes
\begin{equation}
	\label{eqn23} \dot{a}^2 = \frac{2 \beta_0 G}{a^{2(\zeta -1)}} - \lambda a^2,
\end{equation}
where $\beta_0 = \pi \rho_0$. 

The scale factor can be calculated from Eq. (23) and has the form
\begin{equation}
	\label{eqn24} a(t) = \left(\frac{2 \beta_0 G}{\lambda}  \right)^{1/2\zeta} \left[sin(\zeta \sqrt{\lambda}t)\right]^{1/\zeta}.
\end{equation}

This exact form of solution is not necessary here for our purpose. First of all we see that $\dot{a}$ vanishes at $a=\left(\frac{2 \beta_0 G}{\lambda}  \right)^{1/2\zeta}$, after which $\dot{a}$ becomes negative collapses. This happens within a finite time after the beginning of the expansion. From the definition of the particle horizon and Eq. (23), one can find the value of $L_H$ at the turning point.

Hence, one can obtain
\begin{equation}
	\label{eqn25} L_H (turning) = {\frac{1}{(2 \zeta \lambda^{1/2})}}. B{\left(\frac{\zeta -1}{2 \zeta},\frac{1}{2}\right)},
\end{equation}
where $B(p, q)$ is the Euler beta function. 

Then see that at the turning point
\begin{equation}
	\label{eqn26} \frac{\mathbb{S}}{\mathbb{A}} \sim \sigma {\lambda}^{1/(\zeta -1/2)},
\end{equation}
up to factors of unity. 

For $\zeta \geq 1$, the power of $\lambda$ is positive and so the ratio $`{\mathbb{S}}/{\mathbb{A}}'$ is very small at the turning point. We can consider what happen near the final stage of collapse, where the scale factor reaches the Planckian scale. By symmetry $L_H \sim \frac{2 a_0}{a(turning)}{L_H} \sim {\lambda}^{1/(\zeta -1/2)} $ at this time. The scale factor at the Planck time $t = 1$ is $a(t = 1) = \left(\frac{2 \beta_0 G}{\lambda}  \right)^{1/2\zeta} [sin(\zeta \sqrt{\lambda})]^{1/\zeta} $. This ratio is ${\mathbb{S}}/{\mathbb{A}} \sim {\lambda}^{1/\zeta -1/2}$. Hence, for a small $\lambda$ the ratio ${\mathbb{S}}/{\mathbb{A}}\sim {\lambda}^{1/(\zeta -1/2)} >>1$ for $1< \zeta \leq 2$. Therefore, we find that the ratio $S/A$ reaches unity at the same time after the turning point, and the harmonic bound becomes violated thereafter, but still well in a classical phase when the universe is still very large~\cite{Kaloper}.

\section{Observational data analysis}

In this section,  we are going to test the viability of the model using the recent observational data, namely, the observational Hubble data (OHD)~\cite{Ratra,Moresco}. We initially utilize a standard compilation of 30 measurements of Hubble data acquired via the differential age method (DA). This method allows for the estimation of the universe's expansion rate at redshift $z$. The function $H(z)$ is defined as $H(z) = -\frac{dz/dt}{(1+z)}$, applicable within the interval $0.07 < z < 2.0$. In conducting this analysis, we minimize
\begin{eqnarray}
	\chi^2_{H}=  \sum_{i=1}^{30} \frac{[H_{Th}-H_{Obs}(z_i)]^2}{\sigma^2_{H(z_i)}},
\end{eqnarray}
where $H_{Th}$, $H_{Obs}$ and $\sigma_{H(z_i)}$ denotes the theoretical value, observed value and stranded error of $H$ respectively for the Table of 30 points of $H(z)$ data.

\begin{figure}[htbp]
	\centering
	\includegraphics[scale=0.20]{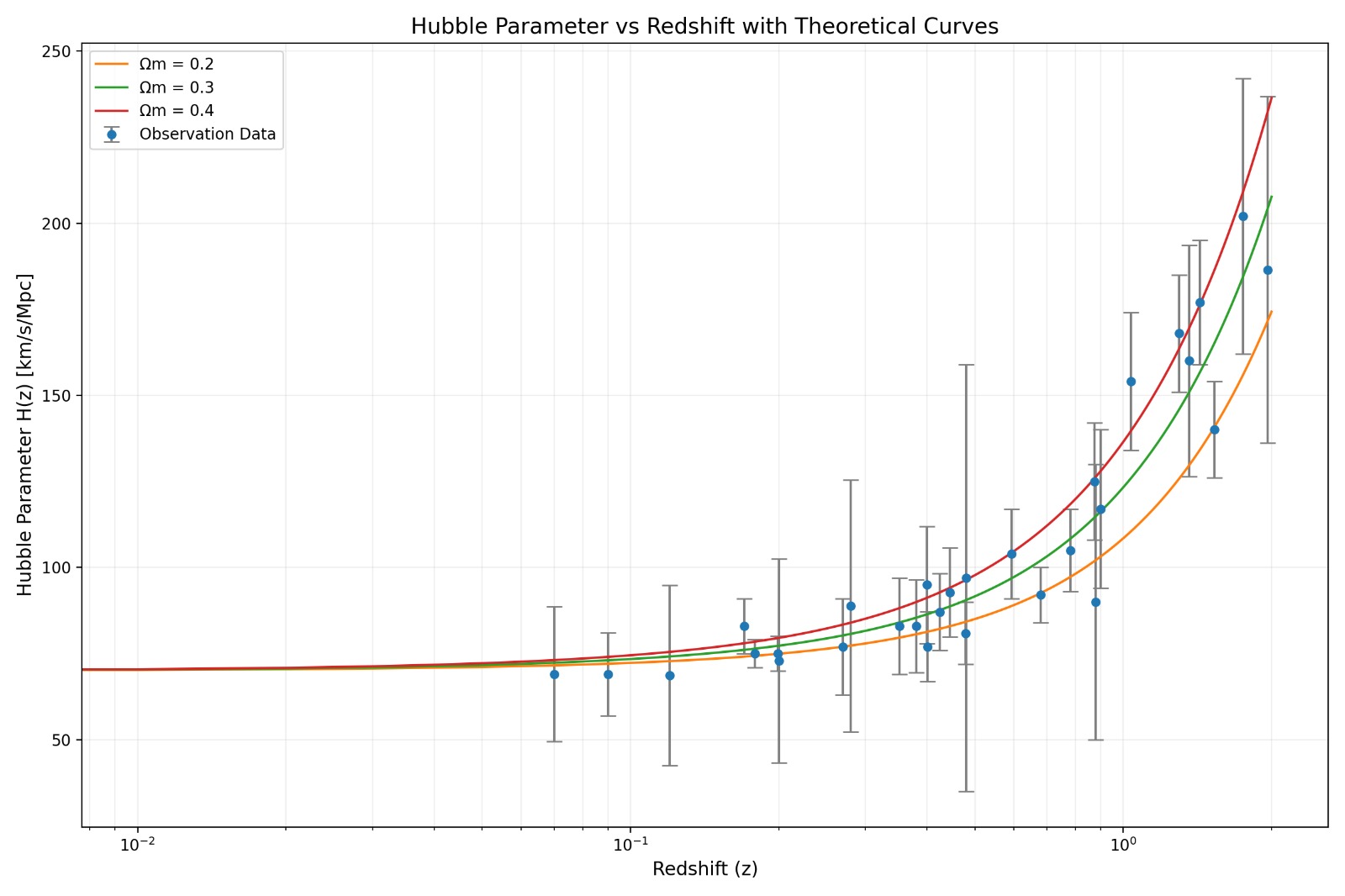}
	\caption{ The plot of Hubble parameter $H(z)$ vs redshift $z$ for the theoretical curves for different values of matter density parameter $\Omega_m = 0.2, 0.3, 0.4$ and the blue dots depict the 30 points of the Hubble observational dataset.}
\end{figure}

\begin{figure}[htbp]
	\centering
	\includegraphics[scale=0.20]{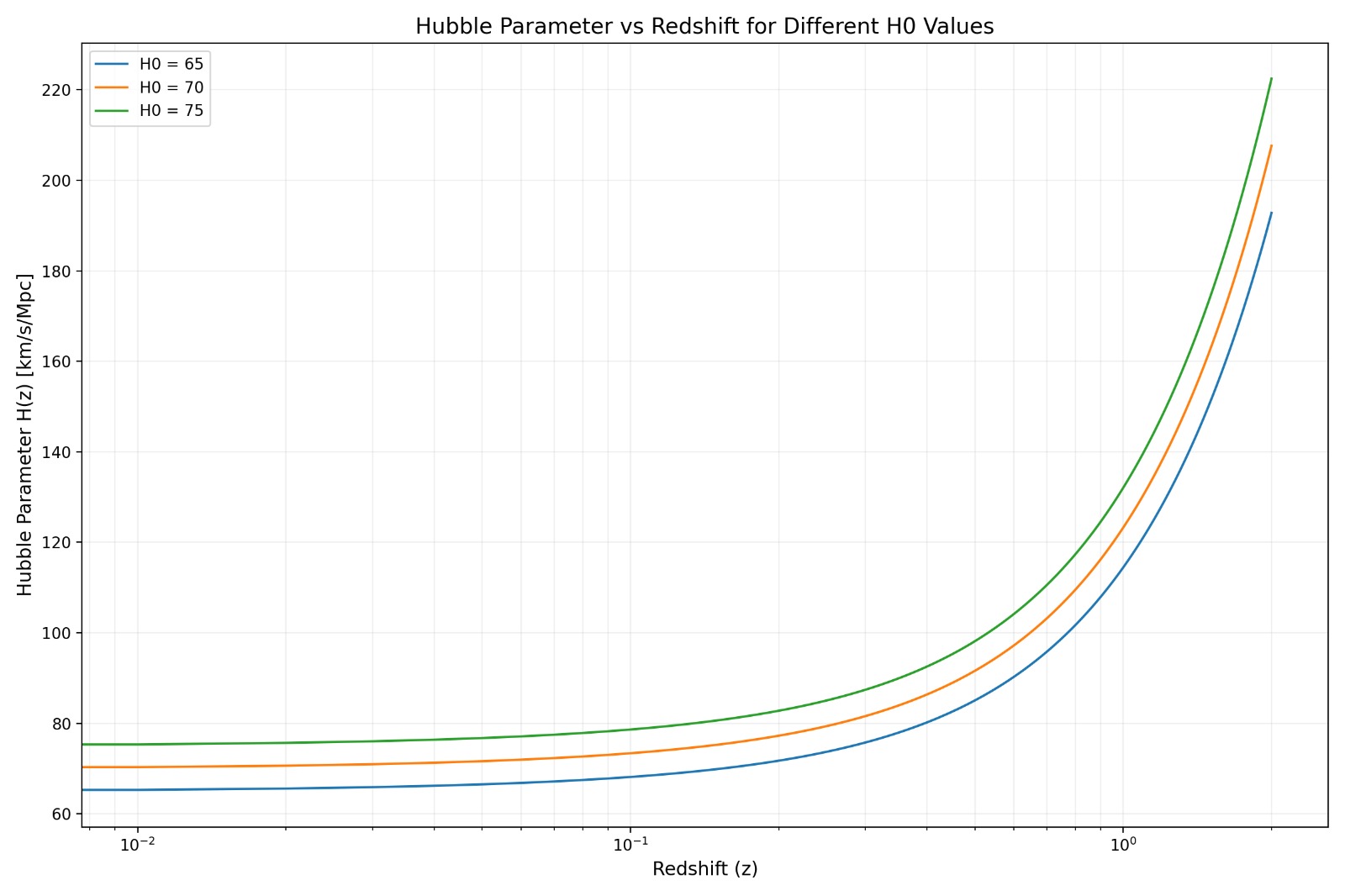}
	\caption{The plot of Hubble parameter $H(z)$ vs redshift $z$ for different values of $H_0 = 65,70, 75$ (in the unit $km/s/Mpc$).}
\end{figure}

\begin{figure}[htbp]
	\centering
	\includegraphics[scale=0.20]{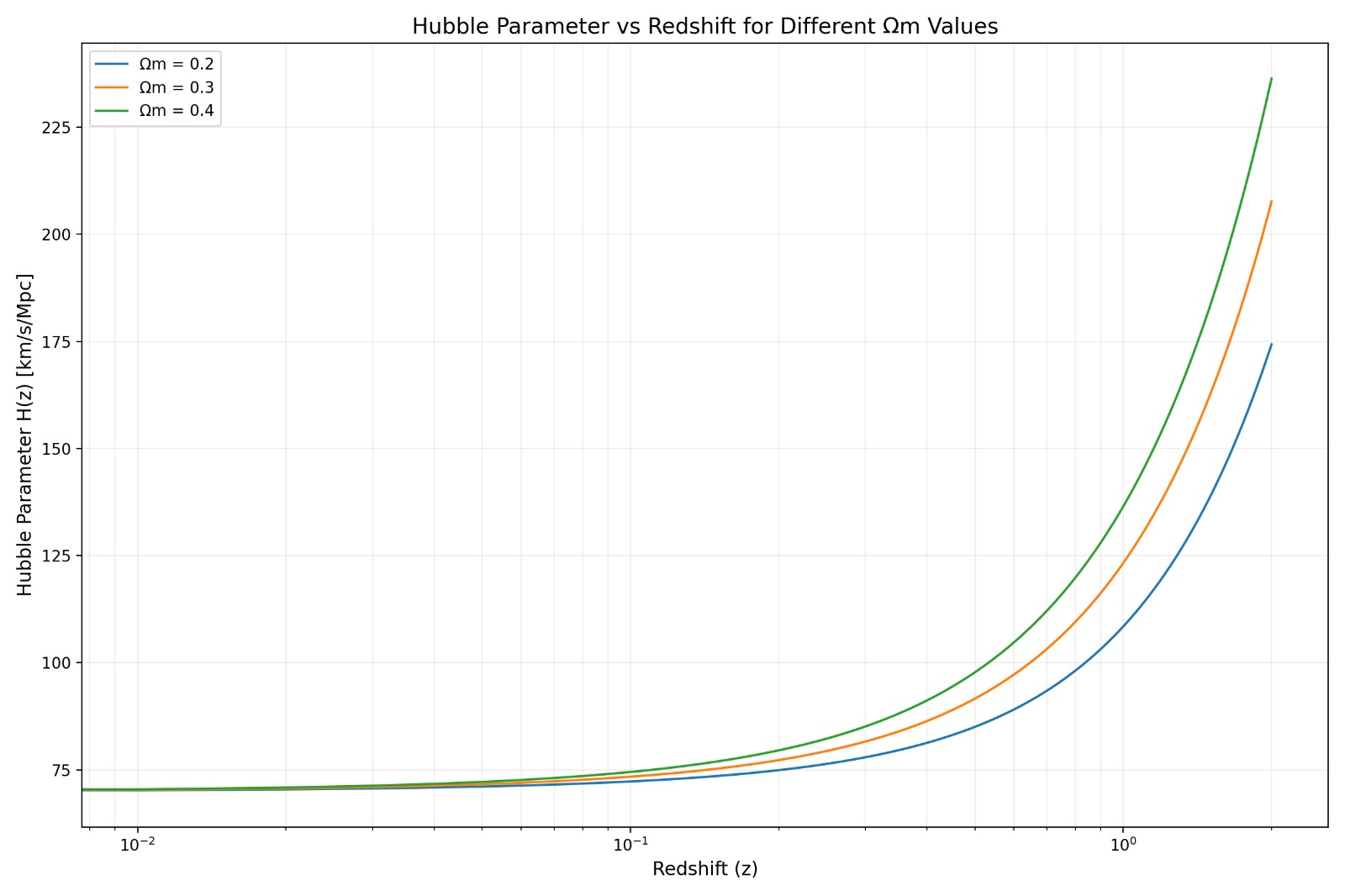}
	\caption{The plot between Hubble parameter $H(z)$ versus redshift $z$ for different values of matter density parameter $\Omega_m = 0.2, 0.3, 0.4$.}
\end{figure}

\begin{figure}[htbp]
	\centering
	\includegraphics[scale=0.20]{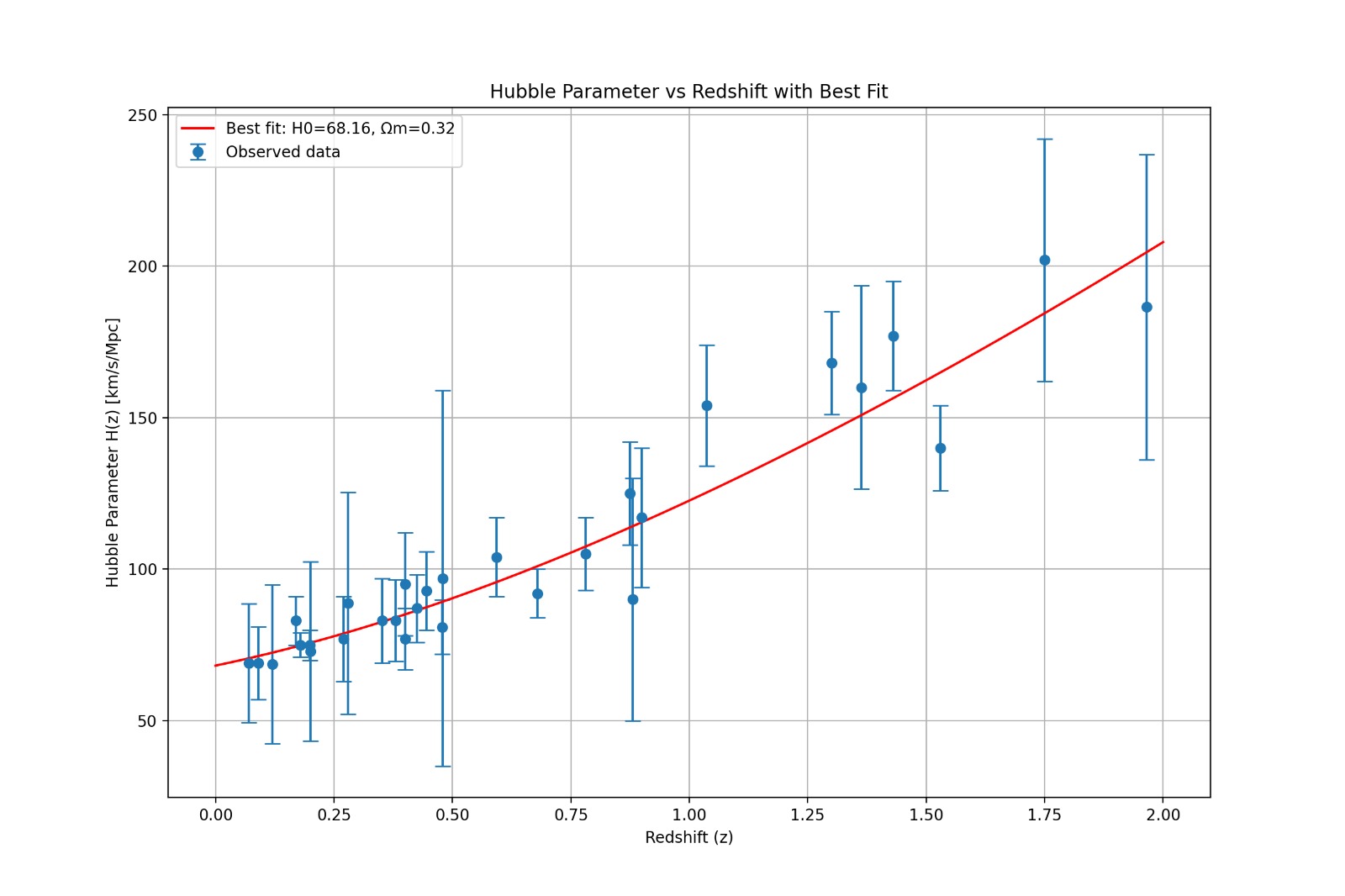}
	\caption{The plot for Best fit of Hubble parameter $H(z)$ versus redshift $z$ with the values of $H_0 =68.16$ and $\Omega_m=0.32$ for the theoretical model (red curve) and the blue dots depict the 30 points of the Hubble data.}
\end{figure}

\begin{figure}[htbp]
	\centering
	\includegraphics[scale=0.20]{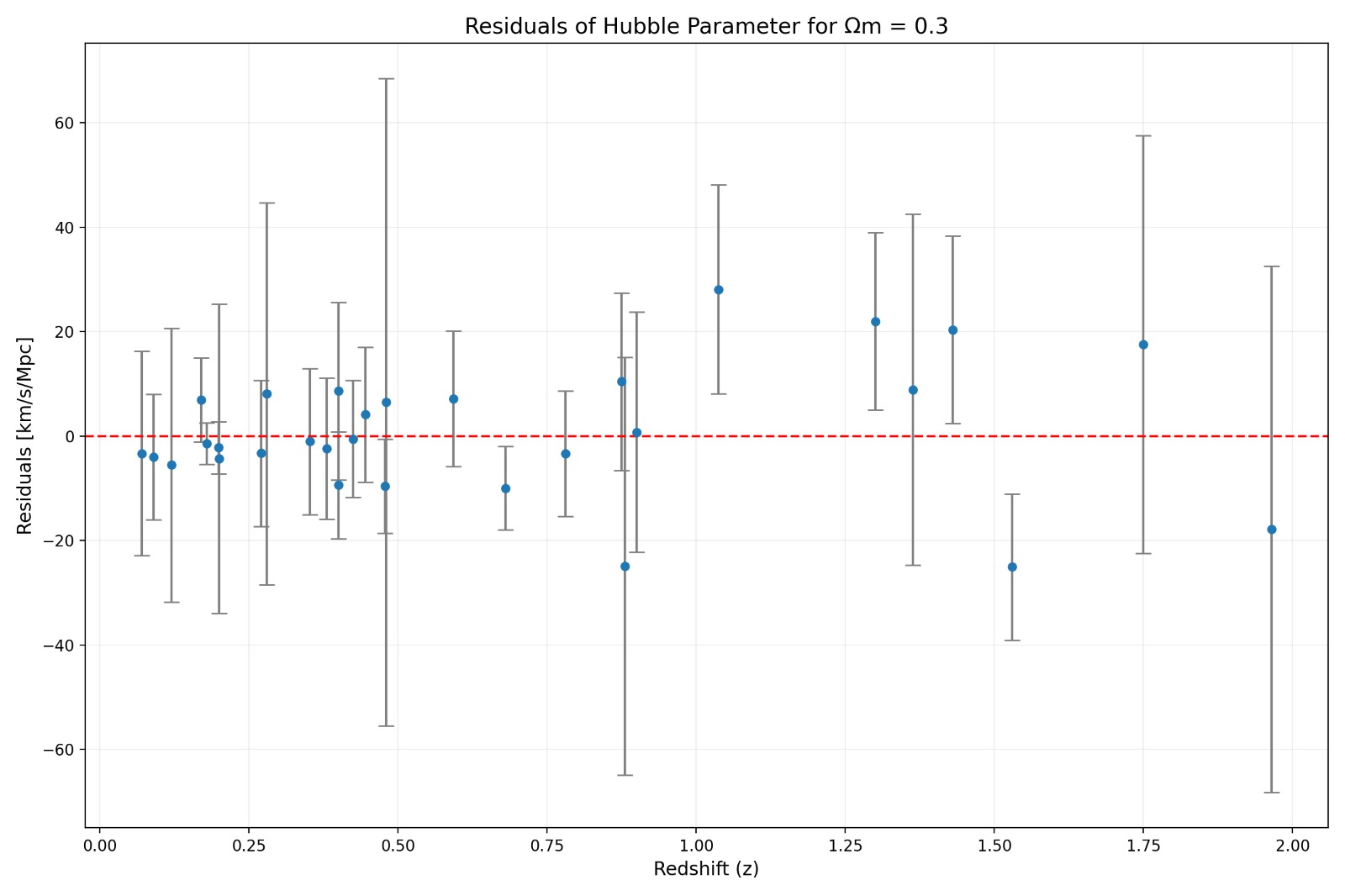}
	\caption{The plot of Residuals of Hubble parameter versus redshift $z$ for the theoretical model (red dotted curve) for the value of $\Omega_m=0.3$ and the blue dots depict the 30 points of the Hubble data.}
\end{figure}

\begin{figure}[htbp]
	\centering
	\includegraphics[scale=0.20]{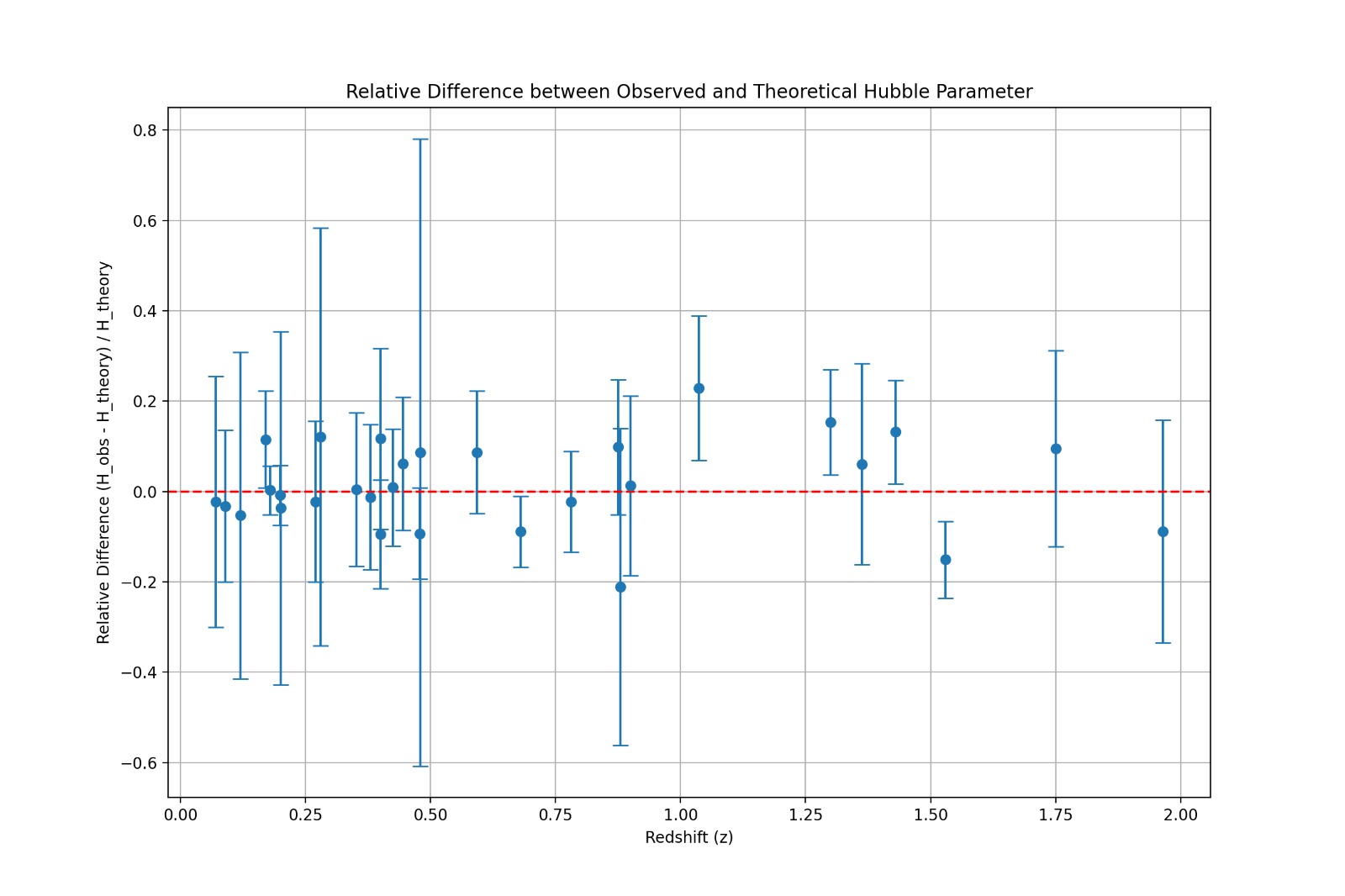}
	\caption{The plot of relative difference between observed and theoretical Hubble parameter versus redshift $z$ and the blue dots depict the 30 points of the Hubble data.}
\end{figure}

\begin{figure}[htbp]
	\centering
	\includegraphics[scale=0.20]{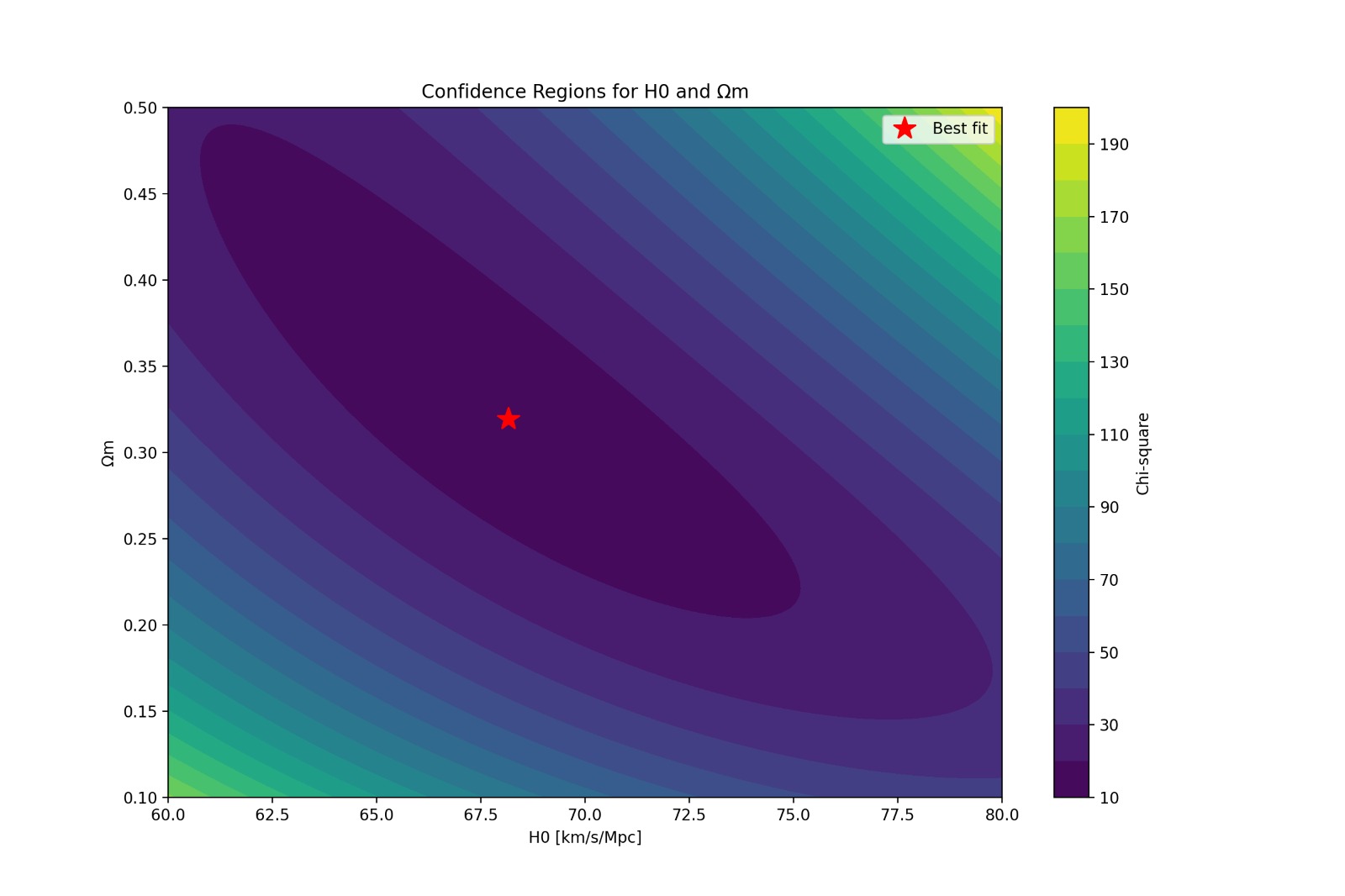}
	\caption{The contour plot for the confidence regions for $H_0$ and $\Omega_m$ with the Chi-square scaling.}
\end{figure}

\section{Physical justification of the graphical plots}

Figure 1 indicates that the $\Lambda CDM$ model with $\Omega_m=0.3$, is a good match to the current observational data of the Hubble parameter. The green curve ($\Omega_m=0.3$) aligns best overall with the observational data, especially in the intermediate redshift range $0.1<z<1.5$. Hubble data alone provides supportive but not definitive constraints due to measurement uncertainties. Figure 2 shows how the Hubble parameter $H(z)$ evolves with redshift $z$, for different values of the present-day Hubble constant $H_0$. Here we isolate the impact of varying $H_0$ (the expansion rate of the universe today) on the predicted evolution of $H(z)$, while keeping the matter density $\Omega_m$ and cosmological constant $\lambda$ fixed. It is observed that at low-redshift regimes the curves diverge most noticeably near $z \approx 0$, where $H(z)\approx H_0$, so the distinction between values is much clear. However at high-redshift behavior at $z \gg 1$, the curves begin to converge in shape, but not in value. It is observed in Fig. 3 that $H(z)$ increases with redshift for all values of $\Omega_m$.

However, at low redshift $(z \ll 1)$, the curves converge, which indicates that the value of $\Omega_m$  has less impact on $H(z)$. At higher redshift, $H(z)$ diverges for different values of $\Omega_m$. The curve for $\Omega_m=0.4$ is the steepest, which might indicate faster expansion in the early universe due to high density of matter. The Red curve in Fig.~4 indicates the best-fit theoretical model to the observational data, assuming a flat Lambda Cold Dark Matter ($\Lambda$CDM) cosmology. Our model provides a reasonable match to the observational data across the full redshift range. Here the chosen parameters ($H_0=68.16, \Omega_m=0.32$) suggest a universe dominated by dark energy at late times and are also found to be consistent with Planck CMB results. Figure 5 shows the residuals of the Hubble parameter $H(z)$ as a function of redshift $z$, and is a good diagnostic tool to assess the goodness of fit between the theoretical model and observational data. We observe from Fig. 5, that $\Omega_m=0.3$ does not show a systematic bias, it does not consistently under-predict or over-predict the values of $H(z)$. Hence the residual plot supports the conclusion that a model with $\Omega_m=0.3$, fits the data reasonably well. 

In Fig. 6, the X-axis represents the redshift $z$, which indicates how far back in time we are, where $z_0$ being the present and increasing values representing earlier epochs in the universe. The Y-axis represents the relative difference, where the value of $0$ suggests that the observed and theoretical values of the Hubble parameter are in parity. Positive values indicate the observed value is higher than predicted, while negative values indicate it is lower. The red dashed line is the horizontal line at $0$ and serves as the reference for perfect agreement between observation and theory. The contour plot in Fig. 7 suggests the best-fit cosmology in strong agreement with Planck/$\Lambda$CDM parameters. We find that the local value of {\textcolor{blue}{$H_0 \sim 73~km/s/Mpc$}} falls outside the central confidence region, indicating the persistent tension between the early-universe and local measurements. Also the contours form elliptical shapes, indicating correlated uncertainties. In this case, increasing $H_0$ requires lowering $\Omega_m$ and vice versa to maintain a good fit.

The narrowness of the ellipses suggests that the data is fairly constraining, although there is still significant uncertainty.

\section {Conclusion}

We have explored the holographic principle within $(2 + 1)$-dimensional cosmological models by analyzing the dynamics of a perfect fluid governed by a generalized equation of state, as found in Eq. (7). By investigating the implications of this equation across various spatial curvatures, we observe that for a particular value of the parameter $\zeta$, the holographic principle holds true in all flat and open universe models (i.e., $k = 0, -1$). This result is consistent with the findings in $(3 + 1)$-dimensional cosmologies, thereby indicating that the holographic principle is valid across different dimensions and geometries. These findings also contribute to the broader understanding of the holographic principle and offer a foundation for further exploration of entropy bounds in lower-dimensional gravity. However, in the case of a closed $(2 + 1)$-dimensional universe, the holographic principle cannot be realized. Moreover, for the generalized equation of state, we have solved the $(2 + 1)$-dimensional Friedmann equations and examined the fundamental assumptions of the holographic principle within a flat cosmological model during the radiation-dominated era. Our analysis shows that, similar to the case of closed $(2 + 1)$-dimensional universes, the Fischler–Susskind entropy~\cite{Fischler} bound breaks down in $(2 + 1)$-dimensional cosmological models with a negative cosmological constant, irrespective of values of $k$. This suggests that the presence of a negative cosmological constant imposes severe constraints on the validity of the holographic bound. It is also observed that prior to the point of maximum expansion the holographic constraints holds. However, once after the point where the ratio exceeds unity the holographic bound cannot be further maintained. This result is in agreement with that of $(3 + 1)$-dimensions.

The general cosmological solutions of the $(3 + 1)$-dimensional Einstein equations are much complicated and are likely governed by non-integrable dynamics. In contrast, the structure of the theory in $(2 + 1)$-dimensions is simpler, and offers the possibility of making meaningful progress in finding general solutions in a variety of interesting scenarios. This fact, combined with the perception that quantum field theory may be more naturally formulated in three rather than four dimensions, has provided strong motivation for studying Einstein’s theory within $(2 + 1)$-dimensional spacetime.

For the observational analysis, we utilize a compilation of 30 $H(z)$ measurements to constrain the model parameters. To obtain the best-fit values, we adopt the Monte Carlo Markov Chain (MCMC) method based on the most recently available data. The parameter estimation is carried out through a $\chi^2$-minimization technique, using the $H(z)$ dataset as the observational input in our model.

A statistical analysis has been conducted to determine the best-fit values of the model parameters using the $H(z)$ dataset through the Monte Carlo Markov chain (MCMC) method. We have applied the $\chi^2$ minimization technique and examined various values of the Hubble parameter and the matter density parameter, in order to accurately constrain the model parameters.

As the concluding remark we would like to state that the overall investigating results are physically viable as far as observational constraints are concerned. Therefore, after this preliminary success we may think of a more complex cosmological system such as the effects of inclusion in the background viscous fluid, inhomogeneity, Chaplygin gas etc. under general relativistic platform or even in the modified theories of gravity~\cite{x1,x2,x3,x4,x5,x6}. 

\section*{Funding}
This work was supported by the Deanship of Scientific Research, Vice Presidency for Graduate Studies and Scientific Research, King Faisal University, Saudi Arabia (KFU252735).

\section*{Acknowledgments}
PKD and SR would like to acknowledge Inter-University Centre for Astronomy and Astrophysics (IUCAA), Pune, India for providing them Visiting Associateship under which a part of this work was carried out. SR also gratefully acknowledges the facilities under ICARD, Pune at CCASS, GLA University, Mathura.\\

{\bf Conflict of Interest:} All authors have stated that they have no competing interests.\\

{\bf Data Availability Statement:} The paper does not include any data.\\

\end{document}